# Enhancing Stratified Graph Sampling Algorithms based on Approximate Degree Distribution

Junpeng Zhu[1,2], Hui Li[1,2(✉)], Mei Chen[1,2], Zhenyu Dai[1,2] and Ming Zhu[3]

[1] College of Computer Science and Technology, Guizhou University, Guiyang, P. R. China
[2] Guizhou Engineer Lab of ACMIS, Guizhou University, Guiyang, P. R. China
`jpzhu.gm@gmail.com`, `{cse.HuiLi, gychm, zydai}@gzu.edu.cn`
[3]National Astronomical Observatories, Chinese Academy of Sciences, Beijing, P. R. China
`mz@nao.cas.cn`

**Abstract.** Sampling technique has become one of the recent research focuses in the graph-related fields. Most of the existing graph sampling algorithms tend to sample the high degree or low degree nodes in the complex networks because of the characteristic of scale-free. Scale-free means that degrees of different nodes are subject to a power law distribution. So, there is a significant difference in the degrees between the overall sampling nodes. In this paper, we propose a concept of approximate degree distribution and devise a stratified strategy using it in the complex networks. We also develop two graph sampling algorithms combining the node selection method with the stratified strategy. The experimental results show that our sampling algorithms preserve several properties of different graphs and behave more accurately than other algorithms. Further, we prove the proposed algorithms are superior to the off-the-shelf algorithms in terms of the unbiasedness of the degrees and more efficient than state-of-the-art FFS and ES-i algorithms.

## 1 Introduction

Networks arise as a natural representation of data in various domains, such as social networks, biological and information networks. However, the real-world networks are massive, which makes traditional analytical approaches infeasible. The methods of data reduction [1] are indispensable. There are various methods have been proposed to cope with the challenge of data reduction, ranging from the principal components analysis to clustering analysis to sampling. Because sampling is an efficient and critical technique for solving massive data analysis bottlenecks, in this paper, we focus on researching graph sampling technique to accelerate the large network data analysis.

Conceptually, we could divide the graph sampling algorithms into three groups: node, edge, and topology-based sampling. In the complex networks, because of the characteristic of scale-free [2], the degrees of different nodes are subject to a power law distribution, the existing algorithms based on the node selection method often over-sample the low degree nodes [3,4,5,7,8,9]. The number of edges corresponding to the high degree nodes is greater, resulting in objectivity the existing algorithms based on the edge selection method tend to draw the high degree nodes [3,4,5,7,8,9]. Simultaneously, the algorithms based on the exploration techniques also have a biased problem to the high degree nodes [3,5,6,7,8,9,10], which often results in the lower accuracy of sampling results.

There are several major challenges in the graph sampling. First, how to smooth the sampling bias between the high degree and low degree nodes in a sampling procedure? To address the issue, we proposed a concept of approximate degree distribution of nodes and devised a stratified strategy using the concept. Second, how to get the approximate degree distribution of nodes? Our experimental results showed that vector clustering algorithms [1] could tackle the problem. And we also found the approximate degree distribution results which are got by k-means are faster and more accurate. Subsequently, we also developed two stratified graph sampling algorithms combining the stratified strategy with the node selection method. Our major technical contributions are the following:

- We proposed the concept of approximate degree distribution (Section 4.2) in the complex networks, and we also devised a stratified strategy using the approximate degree distribution which is got by clustering algorithms. Further, we investigated the number of strata and presented the most superiority empirical value.
- We developed two new graph sampling algorithms combining our stratified strategy with the node selection method (NS). The experimental results showed that our proposed algorithms are more effective and achieved higher accuracy of the sampling results compared with others algorithms.

The rest of this paper is organized as follows. Section 2 presents the proposed algorithms in this paper. Mathematical analysis is in Section 3. Section 4 provides and analyzes the experimental results, and concluded remarks and future work are in Section 5.

## 2  Proposed Algorithms

Allow for the characteristic of scale-free in the complex networks, the difference between nodes is huge. So, we use the stratified sampling technique to smooth sampling error in this paper. It has been proved [5] the algorithms based on the topological structure are difficult to extend from static to streaming graphs. This is the main reason our proposed algorithms are based on the node selection sampling methods (NS). We briefly outline two algorithms and give a formal description in this section.

### 2.1  NS-d Algorithm

We propose Node Degree-Distribution Sampling (for brevity NS-d) algorithm based on the node selection method (NS [9]) and stratified strategy of the nodes. We give the sampling steps of the NS-d in ALGORITHM 1. The algorithm consists of three steps. First (lines 1-3), NS-d divides the nodes into three strata by k-means. We will explain the reason the number of clusters is 3 in Section 4.2. Second (lines 4-7), NS-d selects the nodes from the set of $N_{high\text{-}degree}$, $N_{medium\text{-}degree}$ and $N_{low\text{-}degree}$ using NS by sampling fraction $\varphi$. Subsequently (line 7), NS-d gets a set of nodes $V_s$ which joins the results of stratified sampling. Finally (line 8), using the graph induce to get a set of edges $E_s$.

---
**ALGORITHM 1**: NS-d ($\varphi$, N, S, D)

**Input:** Sample fraction $\varphi$, Node set N, Edge set S, Degree set D
**Output:** Sample Subgraph S = ($V_s$, $E_s$) // $V_s \square$ N and $E_s \square$ S
1. $V_s$ = null set, $E_s$ = null set
2. k-means = k-means (distanceFunction,3)
3. ($N_{high\text{-}degree}$, $N_{medium\text{-}degree}$, $N_{low\text{-}degree}$) = k-means.run(N, D)
4. $V_{high\text{-}degree}$ = NS($\varphi$, $N_{high\text{-}degree}$)   //NS: Node Sampling Algorithm
5. $V_{medium\text{-}degree}$ = NS($\varphi$, $N_{medium\text{-}degree}$)
6. $V_{low\text{-}degree}$ = NS($\varphi$, $N_{low\text{-}degree}$)
7. $V_s$ = $V_{high\text{-}degree}$ ∪ $V_{medium\text{-}degree}$ ∪ $V_{low\text{-}degree}$
8. Get $E_s$(sourceNode, targetNode) and sourceNode ∈ $V_s$ and targetNode ∈ $V_s$
---

NS-d adds k-means step based on NS algorithm, and the time complexity of the k-means method is `O(kt|N|)`, where k is the number of clusters, t is iterations, and there is `k≪|N|` and `t≪|N|` for the k-means procedure. Therefore, the time complexity of NS-d is `O(kt|N|)+O(|N|+|E|)`, where |N| is the number of nodes, |E| is the number of edges. The space complexity of NS-d is `O(|N|+|E|)`.

### 2.2 NS-d+ Algorithm

NS-d+ is similar in spirit to the NS-d described in Section 2.1. We summarized the sampling steps of NS-d+ in ALGORITHM 2. Specifically (lines 4-7), NS-d+ sorts all nodes from the set of high degree nodes using counting sort [11] algorithm by descending and draws them into the sample by sample fraction $\varphi$.

---
**ALGORITHM 2**: NS-d+ ($\varphi$, N, S, D)

**Input:** Sample fraction $\varphi$, Node set N, Edge set S, Degree set D
**Output:** Sample Subgraph S = ($V_s$, $E_s$) // $V_s \square$ N and $E_s \square$ S
1. $V_s$ = null set, $E_s$ = null set
2. k-means = k-means (distanceFunction,3)
3. ($N_{high\text{-}degree}$, $N_{medium\text{-}degree}$, $N_{low\text{-}degree}$) = k-means.run(N, D)
4. CountingSort($N_{high\text{-}degree}$) → Queue   // degrees ∈ [0,$d_{max}$] and $d_{max}$=O(n), so, using counting sort
5. While i < |$N_{high\text{-}degree}$| *$\varphi$
6.     Queue.get()→ $V_{high\text{-}degree}$
7.     i++
8. $V_{medium\text{-}degree}$ = NS($\varphi$, $N_{medium\text{-}degree}$) //NS: Node Sampling Algorithm
9. $V_{low\text{-}degree}$ = NS($\varphi$, $N_{low\text{-}degree}$)
10. $V_s$ = $V_{high\text{-}degree}$ ∪ $V_{medium\text{-}degree}$ ∪ $V_{low\text{-}degree}$
11. Get $E_s$(sourceNode, targetNode) and sourceNode ∈ $V_s$ and targetNode ∈ $V_s$
---

Because the NS-d+ adds counting sort [11] except for k-means step, and the counting sort is stable and its time complexity is `θ(|N|)`. The time complexity of NS-d+ is `θ(|N|)+O(kt|N|)+O(|N|+|E|)`. The space complexity of the NS-d+ is also `O(|N|+|E|)`. Obviously, the time complexity of these two algorithms is linear time, and the space complexity of these compared with NS algorithm remains unchanged.

## 3 Mathematical Analysis

### 3.1 Analysis of the Validity of Algorithms

Let us suppose, the number of the nodes in the original graph is `|N|`. These algorithms divide the population into `i` strata that are recorded as `N1, N2, N3..., and Ni` respectively. These strata satisfy the following conditions:

$$N_p \cap N_q = \emptyset \text{ and } p, q \in [1, i] \text{ and } p \neq q \quad (1)$$
$$N_1 \cup N_2 \cup ... \cup N_i = N$$

According (1), these strata are no overlapping, together they compose all the population. So,

$$|N| = |N_1| + |N_2| + ... + |N_i| \quad (2)$$

Further,

$$|N| * \varphi = (|N_1| + |N_2| + ... + |N_i|) * \varphi \quad (3)$$
$$= |N_1| * \varphi + |N_2| * \varphi + ... + |N_i| * \varphi \quad (4)$$

Where `|N| * φ` is the sampling results of the node selection method (NS). Equation (4) is the sampling results of our proposed algorithms. It is the fact to these results are equivalent, which proves the validity of our proposed algorithms.

### 3.2 Analysis of the Unbiasedness of Degrees

We explain the unbiasedness of proposed algorithms in this section. For the mean of the population, the estimate used in stratified sampling is:

$$\bar{y}_{st} = \frac{\sum_{k=1}^{i} |N_k| \bar{y}_k}{|N|} = \sum_{k=1}^{i} W_k \bar{y}_k \text{ and } W_k = \frac{|N_k|}{|N|} \quad (5)$$

The $\bar{y}_{st}$ represents the estimate of the mean for the population. The $\bar{y}_k$ represents the estimated value of mean at each stratum.

**Theorem 1.** If in every stratum the sample estimate $\bar{y}_k$ is unbiased, then $\bar{y}_{st}$ is an unbiased estimate of the population mean $\bar{Y}$.

**Proof.** Our proposed algorithms divide nodes into three groups based on the approximate degree distribution of nodes. It results in a small difference of the inner stratum and the difference of the between different strata is large in the set of degree (it is proved in Section 4.2). So, the estimate $\bar{y}_k$ is unbiased for the set of the degree in the individual strata. So,

$$E(\bar{y}_{st}) = E \sum_{k=1}^{i} W_k \bar{y}_k = \sum_{k=1}^{i} W_k \bar{Y}_k \quad (6)$$

Where $\bar{Y}_k$ represents the mean at each stratum. Since the estimates are unbiased in the individual strata. So, the population mean $\bar{Y}$ may be written as:

$$\overline{Y} = \frac{\sum_{k=1}^{i}\sum_{l=1}^{i} y_{kl}}{|N|} = \frac{\sum_{k=1}^{i} |N_k| \overline{Y_k}}{|N|} = \sum_{k=1}^{i} W_k \overline{Y_k} \qquad (7)$$

The key is the variance of $\overline{y}_{st}$ depends only on the variances of the individual stratum means $\overline{y}_k$.

## 4 Experiments

In this section, we explain the reasons our proposed algorithms use k-means and the number of clusters is 3 by experiments first. Next, we empirically compare the performance of our proposed algorithms with other methods, such as NS [9], ES [9], ES-i [5], and FFS [9]. We apply to the full network and sample subgraphs over a range of sampling fraction φ=[5%,25%], and the interval number is 5%. For each sampling fraction, we repeated at least 200 experiments to ensure the accuracy of the sampling results. The following section describes in details.

### 4.1 Datasets

We used three data sets from SNAP [12] in the experiments, namely the Facebook that consists of friends' lists, the Condmat that consists of collaboration network from the arXiv and the Amazon that network was collected by crawling the Amazon website. Table 1 summarizes the global statistics of these three real-world datasets.

**Table 1.** Characteristics of Networks Datasets

| Graph | Nodes | Edges | Clustering Coefficient | Diameter | Average Density | Average Degree |
|---|---|---|---|---|---|---|
| FACEBOOK | 4039 | 88234 | 0.6055 | 8 | $1.08 \times 10^{-2}$ | 46.6910 |
| CONDMAT | 23133 | 186936 | 0.6334 | 15 | $7.0 \times 10^{-4}$ | 16.1618 |
| AMAZON | 334863 | 925872 | 0.2050 | 47 | $1.65 \times 10^{-5}$ | 5.5299 |

### 4.2 Why do the Algorithms Use k-means?

The essence of the concept of approximate degree distribution is to get a better cut for the degrees of different nodes. We consider that clustering technology is superior to other methods for getting the results of approximate degree distribution.

There are many vector clustering algorithms [1] in data mining, such as k-means, k-medoids, EM, DBSCAN. The time complexity of DBSCAN is higher than the k-means algorithm, and there are two parameters: one is Eps [1], the other is MinPts [1]. The time complexity of EM and k-medoids are also greater than the k-means algorithm. We consider the procedure is simple and efficient for getting approximate degree distribution results, and k-means satisfies these conditions. The distance function adopts correlation distance [13], which shows a better interpretation of clustered data [13]. Distance correlation is a measure of dependence between random vectors ($x_i$ and $x_j$) and is defined as follows:

$$d_{ij} = 1 - \frac{(x_i - \overline{x_i})(x_j - \overline{x_j})'}{\sqrt{(x_i - \overline{x_i})(x_i - \overline{x_i})'}\sqrt{(x_j - \overline{x_j})(x_j - \overline{x_j})'}} \quad (8)$$

Silhouette coefficient [14] is a measure that can be used to study the separation distance between clusters. It is usually used to select the most superiority k when the number of clusters is unknown. The silhouette coefficient is defined as:

$$S(i) = \frac{b(i) - a(i)}{\max\{a(i), b(i)\}} \quad (9)$$

Where i represents any object in the datasets, a(i) represents the average distance of i with all other data within the same cluster (cohesion), b(i) represents the lowest average distance of i to all nodes in any other clusters (isolation). The average silhouette coefficient is defined as:

$$S_{avg} = \frac{\sum_{i=1}^{|N|} S(i)}{|N|} \quad (10)$$

The greater the average silhouette coefficient is, the more excellent the number of clusters k is. In this paper, the experimental results show that k is 2 or 3 outperforms [15,16] other values in the Fig.1 (a). When k is 2 in the Fig.1 (b), it has not been better separated to the high degree nodes which are significant and small in Table 2. It is superior to others in the Fig.1 (b) for clustering results when k is 3, and the average silhouette coefficient is also close to the optimal value in the Fig.1 (a). Therefore, it is the most excellent empirical value to that k is 3.

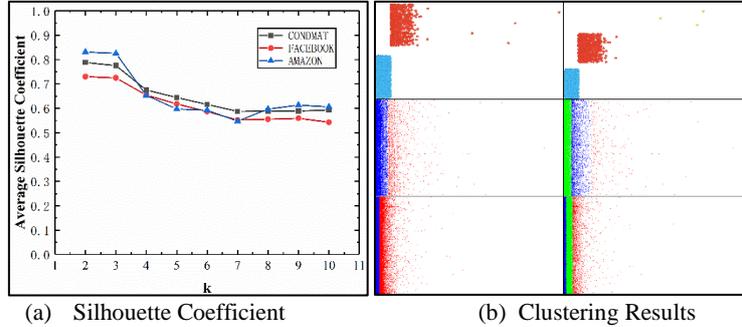

(a) Silhouette Coefficient      (b) Clustering Results

**Fig. 1.** (a) Illustration the relation of k and average silhouette coefficient. (b) Illustration clustering results using k-means when k is 2 and 3 on these datasets in turn.

### 4.3 Accuracy Analysis of Topological Properties

In this section, we compare our proposed algorithms with state-of-the-art algorithms of three groups, which are NS, ES, ES-i, and FFS separately on the degree (distribution), clustering coefficient (distribution), density, and diameter.

**Table 2.** The Percentage of Nodes of Different Clusters

| Graph | Low degree | Medium degree | High degree |
|---|---|---|---|
| FACEBOOK | 91% | 9% | 0% (extremely tiny) |
| CONDMAT | 81% | 18% | 1% (approximate) |
| AMAZON | 83% | 16% | 1% (approximate) |

We acquire that ES-i [5] outperforms ES by experiments in the Fig. 2, Fig. 3, Fig. 4 and Fig. 5. This is mainly due to the concept of graph induce [5], which shows that it is effective to the concept. It is well-known that high degree nodes are often the hubs nodes [3,5], which serve as good navigators through the graph.

The density of NS-d and NS-d+ is greater than NS, and NS-d+ is superior to other methods in the Fig. 2. There are three primary reasons as follows: one is the stratified strategy; two is the concept of graph induce; three is high degree nodes. The density is proportional to the real number of edges according to its definition. So, our proposed stratified strategy enhances the connectivity of sample graphs. It is significant to balance the sampling proportion between high degree and low degree nodes.

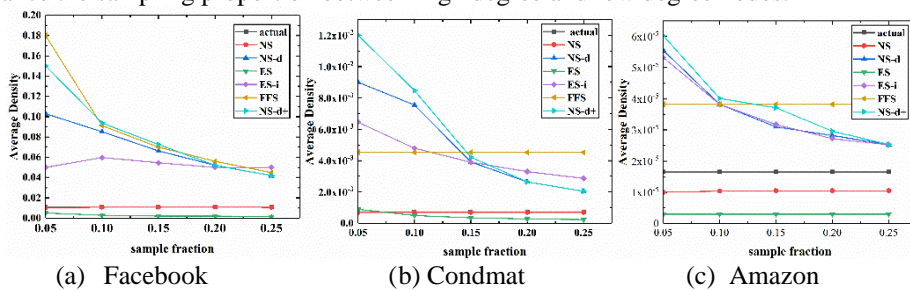

(a) Facebook    (b) Condmat    (c) Amazon

**Fig. 2.** Illustration of average density on the three real-world datasets.

The diameter of NS-d and NS-d+ is close to original graphs in the Fig. 3. The diameter is the maximum of the shortest path. In the graphs, many shortest paths usually pass through these high degree nodes. This is mainly due to that our stratified strategy achieves a balance between the low degree nodes and high degree nodes, which smooth the biasedness of NS that is likely to sample low degree nodes.

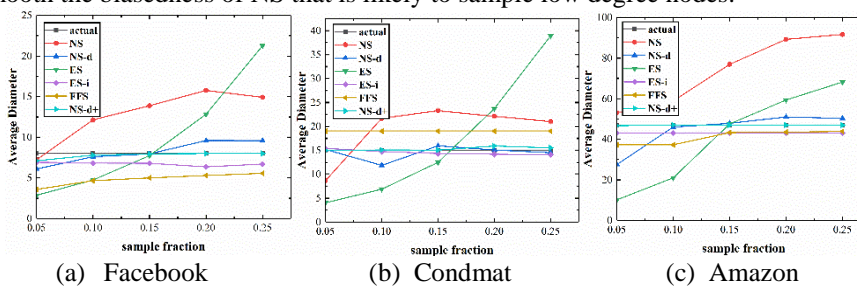

(a) Facebook    (b) Condmat    (c) Amazon

**Fig. 3.** Illustration of average diameter on the three real-world datasets.

Next, we compare the average degree and degree distribution of 15% on three real-world datasets by experiments in the Fig. 4. The experimental results show NS-d and

NS-d+ outperform the other algorithms. This is also mainly due to the stratified strategy which smooth the biasedness of degrees. And the results of NS-d+ are superior to NS-d. The primary reason is the greater the degrees of different nodes are, the more crucial the nodes are. NS-d+ sorts all nodes and draws them from higher to lower by φ, which results in that NS-d+ is more accuracy than NS-d.

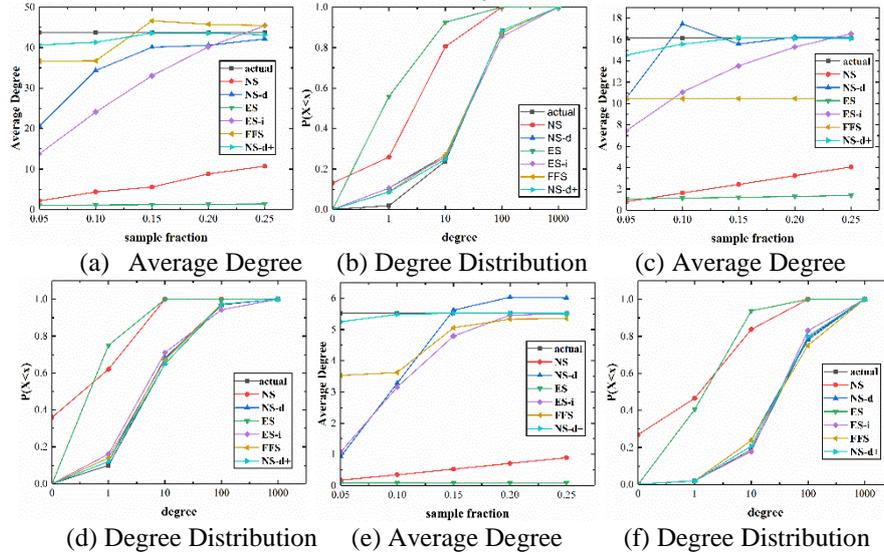

(a) Average Degree  (b) Degree Distribution  (c) Average Degree

(d) Degree Distribution  (e) Average Degree  (f) Degree Distribution

**Fig. 4.** Illustration of average degree, degree distribution on three real-world datasets in turn.

Finally, Fig. 5 shows the average clustering coefficient and clustering coefficient distribution of 15% on the three datasets. We also acquire that NS-d and NS-d+ outperform the other methods, leading to these algorithms could capture the transitivity of the graph because of the stratified strategy of nodes.

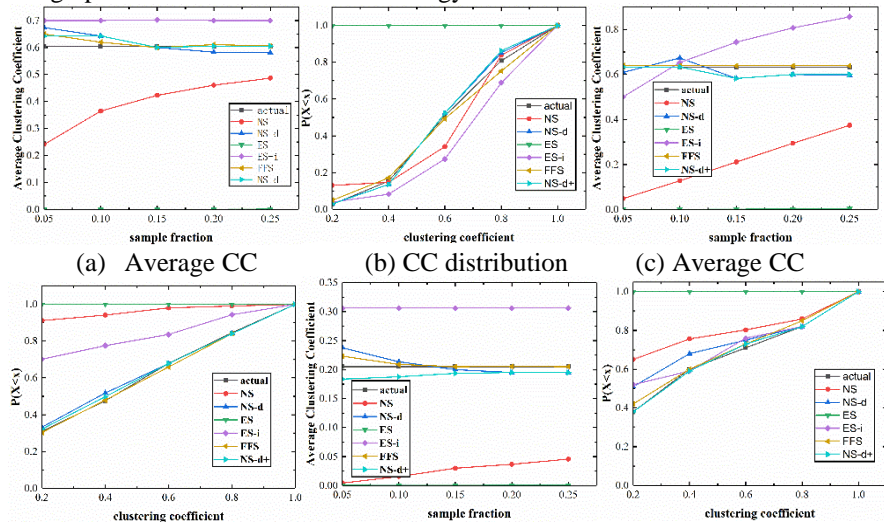

(a) Average CC  (b) CC distribution  (c) Average CC

(d) CC Distribution   (e) Average CC   (f) CC Distribution

**Fig. 5.** Illustration of average clustering coefficient and clustering coefficient distribution on the three real-world datasets in turn.

### 4.4  Comparison of Runtime with state-of-the-art Algorithms

We compare the NS-d and NS-d+ with FFS and ES-i in terms of the average runtime in the Fig. 6. Obviously, the average runtime of NS-d and NS-d+ are both much lower than FFS algorithm. The FFS is a BFS-based algorithm and requires many passes over the edges in a sampling procedure, which results in tangible the time complexity of FFS is polynomial time. ES-i requires two passes over the edges. The time of passing over edges is usually far greater than nodes because there is a huge difference between them on the majority real-world datasets. The NS-d adds k-means step, and NS-d+ algorithm adds k-means and sort steps. The time complexity of k-means is `O(kt|N|)`, and the time complexity of counting sort is `Θ(|N|)`. And NS-d and NS-d+ only require one pass over edges. So, the time complexity of our proposed algorithms is lower than FFS and ES-i.

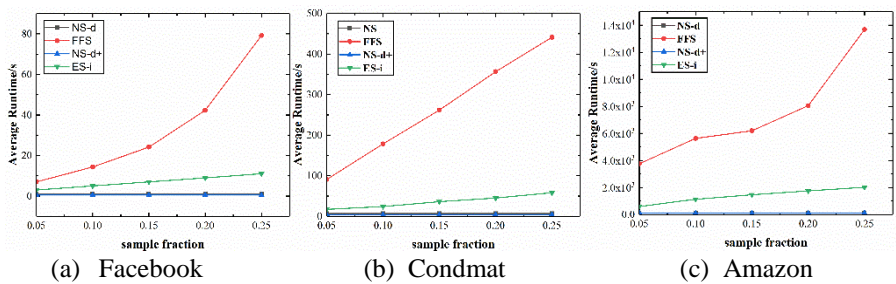

(a) Facebook   (b) Condmat   (c) Amazon

**Fig. 6.** Illustration average runtime on the three real-world datasets in turn.

## 5  Conclusion and Future Work

In this paper, we proposed the concept of approximate degree distribution and devised a stratified strategy using it. Further, we investigated the number of strata and presented the most superiority empirical value of k. Subsequently, we developed two new graph sampling algorithms combining the proposed stratified strategy with the node selection method (NS). The experimental results showed that these algorithms not only accurately preserve more topological properties than state-of-the-art algorithms, but also prove the effectiveness of the approximate degree distribution strategy. Meanwhile, our proposed algorithms outperformed the state-of-the-art FFS and ES-i in terms of average runtime.

Much work still remains. We have focused exclusively on the stratified sampling algorithms on graphs. More sampling strategies should be considered in the different strata. It also will be interesting to extend stratified graph sampling algorithms from static to streaming graphs.

**Acknowledgements.** This work was supported by the Fund by The National Natural Science Foundation of China (Grant No. 61462012, No. 61562010, No. U1531246), Guizhou University Graduate Innovation Fund (Grant No. 2017078) and the Innovation Team of the Data Analysis and Cloud Service of Guizhou Province (Grant No. [2015]53).